# The guide to the guiding center aka pseudo-momentum operator construction

E.L. Rumyantsev and A.V. Germanenko

School of Natural Science and Mathematics, Ural Federal University,

620002 Ekaterinburg, Russia

**Abstract**

The strictly gauge invariant approach to the construction of the analog of guiding center integrals of motion in spatially homogeneous/inhomogeneous constant magnetic fields is considered. With their help the gauge invariant equations, describing the wave functions of highly degenerate Landau levels in the "classical" non-relativistic case, are formulated. The proposed gauge-invariant approach was used also for the construction of the equations describing the quasi-relativistic carriers' behavior in the homogeneous /inhomogeneous magnetic field in the single layer graphene.

## 1 Introduction

Guiding center approximation (or drift approximation) is a well-known and powerful theoretical tool to describe the "classical" charge particle motion in plasma in a strong magnetic field [1]. This most widely used approach allows to decouple fast helical motion of the particle about a local magnetic line from the slow bounce and drift motions along and across magnetic field lines [2,3]. The notion of the guiding center operator as the certain operator integration constant arises also in the quantum mechanical description of the motion in a constant, spatially uniform magnetic field [4,5,6]. In what follows we intend to propose the gauge invariant method of constructing the so called pseudo-momentum operators which can be used for labelling wave functions of the highly degenerate Landau levels and which are directly connected to the guiding center variables in their classical meaning.

## 2 Uniform magnetic field problem revisited

The motion of a particle of mass $m$ and charge $q$ in uniform constant magnetic field is one of the most studied quantum systems. Due to specific algebraic structure of the Hamiltonians considered, as in relativistic case (Dirac Hamiltonian), so in non-relativistic case (Schrodinger Hamiltonian), the energy spectrum can be easily obtained without turn to the solution of corresponding differen-tial equations. Nevertheless, to our point of view there are some questions to be clarified concerning the derivation of the Eigen wave functions in this seemingly simple and thoroughly scrutinized problem. The problem hinges on the necessity to fix the form of the vector potential to achieve this goal. Starting from the papers published by E. H. Kennard, C. C. Darwin and V. Fock [7,8,9] it was common to use mainly circular gauge $A = [B \times r]/2$. For the beginning, we reconsider this simplest case of non-relativistic 2D motion of the particle with the charge $|q|$ in the X-Y plane perpendicular to uniform constant external magnetic field $B > 0$ directed along Z-axis. To simplify our consideration, we neglect spin. The Hamiltonian to be considered in the first quantization runs as

$$\widehat{H} = \frac{m}{2}(\hat{v}_x^2 + \hat{v}_y^2) \qquad (1)$$

Where $v = (p - |q|A)/m$, due to minimal coupling hypotheses. Hereafter $\hbar = c = 1$. So defined velocity component operators satisfy the following commutation rule $\left[\hat{v}_x^{(+)}, \hat{v}_y^{(+)}\right] = i/l_B^2 m^2$ $l_B = \sqrt{1/|q|B}$ (henceforth we put $\hbar = c = 1$). The index $(+)$ is used to underline that we describe motion of the particle with the charge $|q|$. The velocity operators can be redefined to reveal the equivalence of the considered problem to 1D problem of harmonic oscillator. To this purpose, the "quasi-position" $\widehat{Q} = \hat{v}_x^{(+)} m l_B^2$ and "quasi-momentum" $\widehat{P} = m\hat{v}_y^{(+)}$ operators can be introduced which fulfill the usual commutation rules $[\widehat{Q}, \widehat{P}] = i$ valid for the position and momentum operators. The Hamiltonian in these operators is formally equivalent to the traditional 1D harmonic oscillator one. This redefinition allows also to construct Bose operators $\hat{a} = l_B m(\hat{v}_x^{(+)} + i\hat{v}_y^{(+)})/\sqrt{2}$, $\hat{a}^+ = l_B m(\hat{v}_x^{(+)} - i\hat{v}_y^{(+)})/\sqrt{2}$ subjected to the commutation relation $[a, a^+] = 1$, valid in considered case of constant spatially homogeneous magnetic field. The Hamiltonian (1) in these operators acquires the form $\widehat{H} = \omega_c(\hat{a}^+\hat{a} + 1/2)$ where $\omega_c = qB/m$. Choosing symmetric gauge $A = B(-y, x)/2$, it is possible to introduce an additional pair of Bose operators commuting with $\hat{a}$ and $\hat{a}^+$ by simple changing the sign of the charge and interchanging annihilation/creation operators [5,10]

$$\hat{b}^+ = \frac{l_B m}{\sqrt{2}}\left(\hat{v}_x^{(-)} + i\hat{v}_y^{(-)}\right) = \frac{l_B}{\sqrt{2}}[(p_x + |q|A_x) + i(p_y + |q|A_y)] \qquad (2)$$

$$\hat{b} = \frac{l_B m}{\sqrt{2}}\left(\hat{v}_x^{(-)} - i\hat{v}_y^{(-)}\right) = \frac{l_B}{\sqrt{2}}[(p_x + |q|A_x) - i(p_y + |q|A_y)]$$

It is common to define with the help of these operators the coordinates of the center of circular orbit along which the charged particle is gyrating (guiding center operators) [5]. It must be stressed that so written expressions for $\hat{b}, \hat{b}^+$ are misleading. If we assume that $A_i$ in (2) are really the components of a vector potential, we are to accept that the gauge invariance of our solutions is violated. Really, $\hat{v}_i^{(-)}$ in this case are to be identified with the velocity operators for the particle with the charge $-|q|$ (positron?!) which cannot appear in our non-relativistic theory. Moreover, the straightforward evaluation e.g. of the commutator $[\hat{a}, \hat{b}^+]$ leads to the following condition to be imposed on chosen gauge

$$[\hat{a}, \hat{b}^+] = -il_B^2|q|[\partial_x A_x - \partial_y A_y] + l_B^2|q|[\partial_y A_x + \partial_x A_y] \quad (3)$$

Which is zero for the symmetric gauge $\boldsymbol{A} = B(-y, x)/2$ only. So, we are to write e.g. operator $\hat{b}$ as

$$\hat{b} = \frac{l_B}{\sqrt{2}}[(p_x + |q|\tilde{A}_x) - i(p_y + |q|\tilde{A}_y)] \quad (4)$$

Where $\tilde{A}_i$ are the components of some vector field, determined in "**fixed**" symmetric gauge as $\tilde{A}_i = -A_i$. In order to lend support to this statement let us consider the problem of guiding center operators from another point of view which has been discussed e.g. in [11]. We start from classical description where it is possible to solve the problem of the motion in constant magnetic field employing extra conserved quantity $\boldsymbol{k}$ [12,13]. This vector emerges in classical description when we integrate the equation of the motion $m\,d\boldsymbol{v}/dt = |q|\boldsymbol{v} \times \boldsymbol{B}$ with respect to time, with the result $m\boldsymbol{v} = |q|\boldsymbol{r} \times \boldsymbol{B} + \boldsymbol{k}$. The meaning of the integration constant $\boldsymbol{k} = m\boldsymbol{v} - |q|\boldsymbol{r} \times \boldsymbol{B}$ is clarified after proper scaling and rotation [13,14,15]. The vector $\boldsymbol{R}_0 = [\boldsymbol{k} \times \boldsymbol{B}]/|q|B^2$ in the classical picture defines the center of particle circular motion (guiding center) fixed at the moment of magnetic field switching on. It has to be mentioned that this integral of motion has been established by Gorkov and Dzyaloshinskii in [16], see also [17]. It is easy to verify that in quantum mechanical description $\hat{\boldsymbol{k}}$ (now $\boldsymbol{k}$ becomes an operator) remains also time-invariant, as

$$\frac{\partial \hat{\boldsymbol{k}}}{\partial t} = i[\hat{H}, \hat{\boldsymbol{k}}] = i\left[\frac{m(\hat{v}_x^2 + \hat{v}_y^2)}{2}, m\hat{\boldsymbol{v}} - |q|[\hat{\boldsymbol{r}} \times \boldsymbol{B}]\right] = 0 \quad (5)$$

The introduced $\hat{\boldsymbol{k}}$ operators are subjected to more strict commutation conditions in considered uniform magnetic field case namely $[\hat{k}_i, \hat{v}_j] \equiv 0$ regardless of the specific choice of the vector potential. This property will be of use for us later on while discussing the graphene behavior under the action of the spatially homogeneous/inhomogeneous constant magnetic fields. The explicit forms of these operators run as follows

$$\hat{k}_x = m\hat{v}_x - \frac{y}{l_B^2} \quad \hat{k}_y = m\hat{v}_y + \frac{x}{l_B^2} \quad (6)$$

Pay attention that contrary to the statement in [18], these operators are strictly gauge invariant. Really, the physical meaning of the terms $|q|[\hat{\boldsymbol{r}} \times \boldsymbol{B}]$ after scaling and rotation (as in the case of $\hat{\boldsymbol{k}}$

) is revealed as the particle coordinates operators in disguise, which of course do not respond to the gauge transformations and the velocity operators $\hat{v}_i$ are gauge invariant by the definition. As $\hat{\mathbf{k}}$ has dimension of momentum, this vector constant is known under the term "pseudo-momentum" (the nomenclature used in [18,19]). As components of $\hat{\mathbf{k}}$ do not commute ($[k_y, k_x] = i/l_B^2$) but nevertheless commute separately with the Hamiltonian, it is useful to construct from them the ladder operators

$$\tilde{b} = l_B (k_y + ik_x)/\sqrt{2} \quad \tilde{b}^+ = l_B (k_y - ik_x)/\sqrt{2} \qquad (7)$$

In order to bring them both simultaneously into play for labelling the eigenfunction. These operators are in one-to-one correspondence with the pair of b-operators used in [5,10] and discussed above, which practically in all the papers known to us, are used for the construction of the "center of rotation" operators. We can require the eigenfunction $\Psi_{n,\lambda}(\mathbf{r})$ belonging to the given n'th Landau level to be simultaneously the eigenfunction of $\tilde{b}$

$$\tilde{b}\, \Psi_{n,\lambda}(\mathbf{r}) = \lambda \Psi_{n,\lambda}(\mathbf{r}) \qquad (8)$$

where $\lambda$ arbitrary complex number $\lambda = \lambda_1 + i\lambda_2$ (the celebrated coherent states). Or we can use operator $\tilde{b}^+\tilde{b}$ ($m > 0$)

$$\tilde{b}^+\tilde{b}\Psi_{n,m}(\mathbf{r}) = m\Psi_{n,m}(\mathbf{r}) \qquad (9)$$

As this second variant describe the states which are all gyrating about fixed center $\mathbf{r} = 0$, they contradict our classical picture. We with the authors [5] adhere to the first choice which has simple and physically clear interpretation and coincides with the classical description. The second variant can be of use for the description of the charge particle motion in an axisymmetric magnetic field with straight field lines dependent only on $|\mathbf{r}|$ [20]. One more comment is due. The defined coherent states formed non-orthogonal over complete set for arbitrary $\lambda$. It is known that this set can be reduced to orthogonal complete set on the von Neumann lattice [21]. One more possibility to use both non-commuting pseudo-momentum operators simultaneously arises when we try to impose periodic boundary conditions following [10,22] with the help of the shifting operators $\hat{T}_x = \exp(i\hat{k}_x x)$ and $\hat{T}_y = \exp(i\hat{k}_y L_y)$. $L_x$ and $L_y$ define a parallelogram where the particle resides. The periodic conditions demand that

$$\hat{T}_x \Psi = e^{i\theta_x}\Psi \quad \hat{T}_y \Psi = e^{i\theta_y}\Psi \qquad (10)$$

These conditions can be fulfilled if and only if $[\hat{T}_x, \hat{T}_y] = 0$ which due to $[k_y, k_x] = i/l_B^2$ impose restrictions on the choice of $L_x$ and $L_y$

$$\frac{L_x L_y}{l_B^2} = 2\pi n \qquad (11)$$

Where $n$ is any integer number.

One more essential for the gauge invariance of $\hat{k}_i$-operators property is revealed, if we present them in arbitrary gauge $A$ in the form ($m\hat{v} = \hat{p} - |q|A$)

$$\hat{k} = \hat{p} - \frac{1}{2}|q|[\hat{r} \times B] - |q|\left(A + \frac{1}{2}[\hat{r} \times B]\right) \quad (12)$$

It is easy to verify that the combination in parentheses is curl independent, thus meaning that integral

$$\int_{r_1}^{r_2} \left(A + \frac{1}{2}[\hat{r} \times B]\right) dr \quad (13)$$

Is path independent. Thus, we can associate with this expression the gradient of some function, which can be dubbed as generalized gradient transformation function

$$A + \frac{1}{2}[\hat{r} \times B] = \nabla\varphi \quad (14)$$

This well-known combination has appeared in the famous paper by J. Schwinger [23] presenting derivation of relativistic electron propagator within original essentially gauge invariant method. It has been shown that this method can be applied for computing non-relativistic propagator as well, though unfortunately this method is rarely used in this context [24,25]. This curl vanishing expression appeared in [23] in the relativistic invariant form $A_\mu(x) + F_{\mu\nu}(x)/2$ where $F_{\mu\nu} = \partial_\mu A_\nu - \partial_\nu A_\mu$. It is easy to check that in our 2D non-relativistic case this expression coincides with (13). Inasmuch according to the commonly accepted prescription gauge transformation function $\varphi$ has no any effect upon wave function other than multiplication by phase factor $expiq\varphi$, and correspondently (as it is prescribed) can be ignored, we are left with the expression $\hat{k} = \hat{p} - \frac{1}{2}|q|[\hat{r} \times B]$. The subtle point is that we cannot identify $[\hat{r} \times B]/2$ with the vector potential $-A$, as in this case $\hat{k}$ would be gauge dependent quantity as it was clarified above. The operator $\hat{k} = \hat{p} + |q|A$ so understood is the mechanical moment for the "anti-particle" and according to charge super-selection rule would be acting in orthogonal Hilbert subspace [26,27]. Such difference in the behavior of these two terms of the pseudo-momentum under gauge transformation is reminiscent of the point of view presented in [28,29,30]. The only difference is that the authors propose to consider peculiarities not in $\hat{k}$ transformation but in the redefined vector potential $A = A^{phys} + A^{pure}$. Their basic postulate is that under gauge transformation $U = \exp(iq\chi)$, these two components transform differently. $A^{pure}$ transforms as the full $A$ ($A^{pure} \to A^{pure} + \nabla\chi$), while $A^{phys}$ transforms in the same manner as does the electric field $E$ thus remaining unchanged $A^{phys} \to A^{phys}$ (see [30] and citing within).

The well-studied problem in homogeneous field nevertheless arises two questions. First, if we discard defined above $\nabla\varphi$ in $\hat{k}$ through gauge transformation of the wave function $\Psi(r) = \Phi(r)\exp i|q|\varphi$, the equation for $\Phi(r)$ will contain "fixed" symmetric vector potential $[B \times \hat{r}]/2$ independently of the form of our initially arbitrary chosen potential $A$. Such property of the considered approach resolves long-standing puzzle of the linear Landau gauge $A_1 = B(0,x)$ or $A_2 = B(-y,0)$. Contrary to common statement, if we rely on the "symmetry" considerations and announce the eigenfunctions to be of the form $\varphi(x)\exp(i\gamma y)$ for $A_1$ ($\varphi(y)\exp(i\gamma x)$ for $A_2$), the obtained wave functions in these gauges does not belong to the space of square integrable functions of the symmetric gauge thus contradicting announced gauge invariance. The problem of the mapping states in $A_1$ gauge to the states in $A_2$ gauge is also not so simple and straightforward as has been clarified in [31]. The outlined approach show, that due to the existence of guiding center integral of motion, starting with arbitrary $A$, we arrive at the equations written in the symmetric gauge which lead uniquely to the solutions with the finite norm. It must be noted that the problem of such possible "uniqueness" of the vector potential choice has been discussed albeit from another point of view in [32,33].

## 3 Non-relativistic particle in the inhomogeneous field

Following the approach outlined above, we present below the construction of an analog of the guiding center operator for spatially inhomogeneous magnetic field [34]. We choose e.g. a magnetic field with a constant gradient given by

$$B(r) = B_0 \frac{x}{L}\hat{z} \qquad (15)$$

Where $B_0$ is a constant and $L \equiv |\nabla \ln(B)|^{-1}$ is the constant gradient length scale. This seemingly oversimplified example of spatially inhomogeneous field is nevertheless important for the description of the charge particle motion in the region of magnetic field reversal leading to formation of current sheets along neutral lines [35,36]. The classical variant of this problem has been discussed in [37,38]. Being inspired by the form of guiding center operators in homogeneous case, we proposed that an analog of gauge invariant pseudo-momentum operator (if exists) in such field is also of the form

$$\hat{k} = (m\hat{v} - |q|\widetilde{A}) \qquad (16)$$

From now on we will use only gauge invariant velocity operators $v = (p - |q|A)/m$, so the (+) index will be omitted. It should be reminded that in no way $\widetilde{A}(r)$ in the expression for $\hat{k}$ can be considered as the vector potential and as so it must remain unchanged under gauge transformation. This vector field is to be determined from the condition that $\hat{\boldsymbol{k}}$ (or its components) commute with the Hamiltonian. This condition is a fortiori fulfilled if as discussed above

$$[m\hat{v}_i, \hat{k}_j] \equiv 0 \quad i,j = 1,2 \qquad (17)$$

Taking into account that $[\hat{v}_x, \hat{v}_y] = i|q|B(\boldsymbol{r})/m^2$, these conditions lead to the following set of equations

$$\partial_x \tilde{A}_x = 0 \quad \partial_y \tilde{A}_x = B(\boldsymbol{r}) \qquad (18)$$

$$\partial_x \tilde{A}_y = -B(\boldsymbol{r}) \quad \partial_y \tilde{A}_y = 0$$

Substituting chosen $B(\boldsymbol{r}) = B(x)$, we infer that only $\tilde{A}_y$ component complies with the required conditions and is

$$\tilde{A}_y = -B_0 \frac{x^2}{2L} \qquad (19)$$

By the way, the arising of such constant of motion (conserved quantity) can be inferred from the classical equation of motion:

$$\dot{v}_x = \frac{|q|}{m} B_0 \frac{x}{L} v_y \quad \dot{v}_y = -\frac{|q|}{m} B_0 \frac{x}{L} v_x \qquad (20)$$

Integrating the second equation we obtain

$$v_y = -\frac{|q|}{m} B_0 \frac{x^2}{2L} + \frac{k_y}{m} \qquad (21)$$

It is easy to verify that going to quantum mechanical description the so defined operator $k_y$ is the required additional integral of motion. As we have only one conserved component of pseudo-momentum we are left with no choice but to state that the Eigen functions of the problem $\hat{H}\Psi_E(\boldsymbol{r}) = E\Psi_E(\boldsymbol{r})$ are simultaneously the Eigen functions of found pseudo-momentum component

$$\hat{k}_y \Psi_{E,\lambda}(\boldsymbol{r}) = \left(-i\partial_y - |q|A_y - |q|\tilde{A}_y\right)\Psi_{E,\lambda}(\boldsymbol{r}) = \lambda \Psi_{E,\lambda}(\boldsymbol{r}) \qquad (22)$$

The vector potential $A_y$ in this expression is any "real" vector potential suitable to our problem, which changes under gauge $U(1)$ transformation. Once more we want to call the reader's attention to the specific behavior of $\hat{k}_y$ operator under gauge transformation. Only $A_y$ term in this expression undergoes change with gauge variation. The term $\tilde{A}_y$ is not affected by this operation as it is in essence the function of particle coordinates and thus is not subjected to gauge transformations.

Now we are going to prove that notwithstanding the specific choice of the vector potential, obtained solutions belong to the same Hilbert space. Really, let us choose the vector potential in the Landau-like (linear) gauge $\boldsymbol{A} = B_0(0, x^2/2L)$. In this gauge $\hat{k}_y = -i\partial_y$ and thus the $y$ component of the canonical momentum is conserved in accord with the commonly accepted approach based on the symmetry of the problem [39]. In this case $\Psi_{E,\lambda}(\boldsymbol{r}) = \varphi_{E,\lambda}(x)\exp i\lambda y$, where $\varphi_{E,\lambda}(x)$ satisfies the equation ($l_{B_0}^2 = 1/|q|B_0$)

$$\left[-\partial_x^2 + \left(\lambda - x^2/2Ll_{B_0}^2\right)^2\right]\varphi_{E,\lambda}(x) = 2mE\varphi_{E,\lambda}(x) \qquad (23)$$

Now, let us try the symmetry-like gauge in this problem, using for this purpose the so called Poincare' or multipole gauge (PMG) [40,41,42,43,44]. In its relativistic covariant form PMG potential satisfies condition $x_\mu A^\mu(x) = 0$. In this case it can be expressed as the integral over the electromagnetic field tensor $F_{\mu\lambda}(x) = \partial_\mu A_\lambda(x) - \partial_\lambda A_\mu(x)$ as [45]

$$A_\mu(x) = \int_0^1 du\, ux^\lambda F_{\lambda\mu}(ux) \qquad (24)$$

In the considered by us non-relativistic limit this expression transforms into

$$\boldsymbol{A}(\boldsymbol{r}) = -\boldsymbol{r} \times \int_0^1 du\, u\, B(u\boldsymbol{r})\hat{\boldsymbol{z}} \qquad (25)$$

It is easy to verify that for spatially homogeneous field we obtain well-known potential in symmetric gauge. In chosen magnetic field with constant gradient, according to (25), the chosen "arbitrary" vector potential is

$$\boldsymbol{A}(\boldsymbol{r}) = B_0(-y, x)\frac{x}{3L} \qquad (26)$$

Correspondently, $\hat{k}_y = -i\partial_y + x^2/6Ll_{B_0}^2$ and $\Psi_{E,\lambda}(\boldsymbol{r}) = \varphi_{E,\lambda}(x)\exp\left[iy\left(\lambda - x^2/6Ll_{B_0}^2\right)\right]$. It is easy to verify that $\varphi_{E,\lambda}(x)$ in this expression coincides with the one in (22), as it satisfies the same equation. Thus it is proved that as in the case of the uniform magnetic field, so in our non-uniform problem the requirement on the wave function to be the eigenfunction of the pseudo-momentum leads to the one-to-one, up to the exponential phase factor, mapping of the solutions belonging to the different gauges. It is interesting to compare our quantum mechanical problem with classical solutions in constant gradient field discussed in [46, 47, 36]. Compare the expression for conserved Y component of the canonical momentum $p_y = mv_y + qA_y = const$ (Formula (1) in [36]) which is in one-to-one correspondence with considered guiding center operator $\hat{k}_y$. It must be noted that the quantum variant of this problem is by far more rich in physics as compared to its classical counterpart. The quantum description for $\lambda > 0$ is given by 1D Schrodinger equation (23) with the "celebrated quartic double-well potential" [48] which is omnipresent in different physical and chemistry problems. The quantum solution differs due to the tunneling effect ("instanton" behavior) from classical prediction of existing for $\lambda > 0$ localized "one-sided" gyration [36] which does not cross $B = 0$ line. Thus, as for $\lambda > 0$ so for $\lambda < 0$, the average particle quantum motion is symmetric relative to neutral magnetic line. It must be stressed than despite the apparent differences in the exponential factors of the considered wave functions the $y$ component of the physically meaningful quantity –the current density remains invariant (as it must be!) under the gauge change. Using the definition of the current density in magnetic field [Landau] it is straightforward to show that in the both gauges the current density is

$$J_y = \frac{iq}{2m}[(\partial_y \Psi^*)\Psi - \Psi^*\partial_y\Psi] - \frac{q^2}{m}A_y\Psi^*\Psi = \frac{|q|}{m}\left(\lambda - \frac{x^2}{2Ll_{B_0}^2}\right)|\varphi(x)|^2 \quad (27)$$

Pay attention that according to this expression, $\lambda$ sign alone does not determine the direction of the particle drift in the considered state. For $\lambda > 0$ all depends on the average mean-square value of the particle deviation along $X$ axes $\langle x^2/2Ll_{B_0}^2\rangle$. For $\lambda < \langle x^2/2Ll_{B_0}^2\rangle$ the particle changes the drift direction. This result is in accord with the classical considerations [36] and at the same time reveals peculiar status of $\hat{\mathbf{k}}$ operators. The physical meaning as an observable must be ascribed without doubt to $J_y$, which means that $\hat{k}_y$ plays some auxiliary role and its consideration as observable is under question. Here it is appropriate to remember (see above) that this characteristic emerges in classic picture not as the constant of motion in its accepted meaning but as the constant of integration over time. This suspicion of the strange role of the guiding center operator in our quantum problem is reinforced by the revision of the solution in homogeneous magnetic field discussed above. The states belonging to, e.g., ground Landau level (from which all others n-levels wave functions can be deduced) are given by the equation

$$\hat{a}\Psi_0(\mathbf{r}) = \frac{l_B m}{\sqrt{2}}(\hat{v}_x + i\hat{v}_y)\Psi_0(\mathbf{r}) = \frac{1}{\sqrt{2}}(\hat{a}_x + i\hat{a}_y)\Psi_0(\mathbf{r}) = 0 \quad (28)$$

Where $\hat{a}_i, a_i^+$ are the Bose operators $[\hat{a}_i, a_j^+] = \delta_{ij}$ of the form

$$\hat{a}_i = -il_B\left(\partial_{x_i} + \frac{x_i}{2l_B^2}\right) \quad a_j^+ = -il_B\left(\partial_{x_i} - \frac{x_i}{2l_B^2}\right) \quad (29)$$

The general solution $\Psi_{0,\lambda}(\mathbf{r}) = \varphi_{0,\lambda}(x)\varphi_{0,i\lambda}(y)$ of the equation () is given by the solutions of two equations

$$\hat{a}_x \varphi_{0,\lambda}(x) = \lambda \varphi_{0,\lambda}(x) \quad \hat{a}_y \varphi_{0,i\lambda}(y) = i\lambda \varphi_{0,i\lambda}(y) \qquad (30)$$

Where $\varphi_{0,\lambda}(x), \varphi_{0,i\lambda}(y)$ are corresponding coherent states and $\lambda = \lambda_1 + i\lambda_2$ arbitrary complex number. So we obtain the highly degenerate set of the wave functions belonging to the same Landau level without invoking the guiding center operators. Thus the use of pseudo-momentum operators in this problem is superfluous. They can be used for clarifying the physical meaning of the numbers $\lambda_{1,2}$ The action of the introduced above operator $\tilde{b} = l_B(k_y + ik_x)/\sqrt{2}$ where $\{k_i\}$ are the discussed pseudo-momentum operators upon these functions is

$$\tilde{b}\Psi_{0,\lambda}(\boldsymbol{r}) = \frac{1}{\sqrt{2}}(a_y + ia_x)\Psi_{0,\lambda}(\boldsymbol{r}) = i\lambda\sqrt{2} \qquad (31)$$

So we can state that really the numbers $\lambda_{1,2}$ labelling the eigenfunctions can be interpreted after scaling by $\lambda_B$ as the corresponding $R_{2,1}$-coordinates of the center of the particle gyration. The usefulness of these operators lies in the fact that with them we can construct gauge invariant equations choosing for the start any appropriate gauge as it was clarified above.

## 4 Graphene in the magnetic fields

An additional but no less important example of proposed approach is due to the fact that introduced above gauge invariant pseudo-momentum operators $\hat{k}_x, \hat{k}_y$ (6) remain valid as the motion constants for the description of the low-energy envelope states in the single layer graphene in homogeneous field and defined above $\hat{k}_y$ (20,21) can be used for labeling states in the perpendicular gradient magnetic field $\boldsymbol{B}(\boldsymbol{r}) = B_0 x\,\hat{\boldsymbol{z}}/L$. Due to commutation conditions defined in (5) for homogeneous field and $\hat{k}_y$ commutator (17) in the gradient field case these operators commute with the Dirac-like Hamiltonian describing carriers behavior in graphene within $\boldsymbol{k}\cdot\boldsymbol{p}$ approach and can serve as the corresponding quantum numbers [49,50]. Due to valley degeneracy of graphene Hamiltonian valid for arbitrary perpendicular magnetic field it suffices as it is common to restrict our consideration to one of the valleys (say K valley) described by the Hamiltonian $\hat{H} = v_F(\hat{Q}^+\sigma^+ + \hat{Q}\sigma^-)$ [51]. Here $\hat{Q} = \hat{\pi}_x + i\hat{\pi}_y$, $\sigma^{\pm} = (\sigma_x \pm i\sigma_y)/2$, and $\sigma_i$ are Pauli matrixes. Consider the behavior of the zero-mode states (if existing) described by the first-order partial differential equation

$$(\hat{\pi}_x + i\hat{\pi}_y)\Psi(\boldsymbol{r}) = 0 \qquad (32)$$

Where $\hat{\pi}_i = -i\partial_i - |q|A_i(\boldsymbol{r})$. It is straightforward to show that the sought-for solutions form the set of coherent states being in one-to-one correspondence with the set describing the degenerate lowest Landau level in the "classical" non-relativistic problem [see (28,29,30)]. As discussed above, in the gradient magnetic field $\boldsymbol{B}(\boldsymbol{r}) = B_0\, x\hat{\boldsymbol{z}}/L$ we can choose any appropriate gauge. In deciding on the gauge $\boldsymbol{A} = (0, B_0 x^2/2L)$ we arrive to the simplest form for $\hat{k}_y = -i\partial_y$ (see discussion above). Labeling the Eigen functions $\Psi_\lambda^T(\boldsymbol{r}) = (\varphi_\lambda(x), 0)\exp(i\lambda y)$ by its eigenvalue $\lambda$ we obtain

$$[-i\partial_x + i(\lambda - x^2/2Ll_{B_0}^2)]\varphi_\lambda(x) = 0 \qquad (33)$$

It follows from (33) that $\varphi_\lambda(x) \sim \exp(\lambda x - x^3/6Ll_{B_0}^2)$. Contrary to the behavior in the considered above non relativistic Schrodinger case where a particle can remain localized along neutral line [36] crossing it hither and thither, the zero-mode carriers in K valley escape to $x = -\infty$ thus destroying the current sheet. The general Landau state $(E \neq 0)$ is given by the solution $\Phi(\boldsymbol{r})^T = (\varphi_1(\boldsymbol{r}), \varphi_2(\boldsymbol{r}))$ of the matrix equation

$$(-EI + v_F \hat{Q}^+ \sigma^+ + v_F \hat{Q} \sigma^-)\Phi(\boldsymbol{r}) = 0 \qquad (34)$$

These equations of the first order can be transformed to the equations of the second order by applying to the equation (34) the operator $EI + v_F \hat{Q}^+ \sigma^+ + v_F \hat{Q} \sigma^-$

$$(EI + v_F \hat{Q}^+ \sigma^+ + v_F \hat{Q} \sigma^-)(-EI + v_F \hat{Q}^+ \sigma^+ + v_F \hat{Q} \sigma^-)\Phi(\boldsymbol{r}) = 0 \qquad (35)$$

As a result, we are to solve two Schrodinger-like equations

$$(-E^2 + v_F^2 \hat{Q}^+ \hat{Q})\psi_1(\boldsymbol{r}) = 0 \quad (-E^2 + v_F^2 \hat{Q}\hat{Q}^+)\psi_2(\boldsymbol{r}) = 0 \qquad (36)$$

Which are super-symmetry (SUSY) connected [52], [53] as the solutions $\Psi^T(\boldsymbol{r}) = (\psi_1(\boldsymbol{r}), \psi_2(\boldsymbol{r}))$ are subjected to the condition $\psi_2(\boldsymbol{r}) = v_F \hat{Q} \psi_1(\boldsymbol{r})/E$. As it has been clarified in [54] the arbitrary solutions of these squared Dirac-like equations being of the second order can contain "superfluous" ones which do not satisfy the initial equation of the first order. The remedy is to consider the function

$$\Phi(\boldsymbol{r}) = (EI + v_F \hat{Q}^+ \sigma^+ + v_F \hat{Q} \sigma^-)\Psi(\boldsymbol{r}) \qquad (37)$$

Which is the solution of the first order equation (34) if $\Psi(\boldsymbol{r})$ is the solution of (35). As we are left with the solution of the one Schrodinger-like equation (36), the procedure outlined above for non-relativistic problem can be at once applied for the analysis of carrier spectrum in graphene. For the chosen gradient magnetic field, the explicit form of the corresponding Schrodinger –like operator is

$$v_F^2 \hat{Q}^+ \hat{Q} = v_F^2(\hat{\pi}_x - i\hat{\pi}_y)(\hat{\pi}_x + i\hat{\pi}_y) = v_F^2[\hat{\pi}_x^2 + \hat{\pi}_y^2 - x/Ll_{B_0}^2] \qquad (38)$$

The difference with the non-relativistic case discussed above resides in the linear in $x$ term (compare with (23)). The wave function solutions as in non-relativistic case demonstrate two types depending on the sign of $\lambda$, described above. The only difference with classical result is that they are shifted in the positive direction along $X$ axis. The symmetry is restored when we consider carrier behavior in $K'$ valley. We will not proceed further with the analysis of the wave function solutions and energy spectrum of (35) which will be considered elsewhere, as our task in the presented paper has been to pave gradient invariant road to the formulation of "proper" wave equations with the help of pseudo-momentum aka guiding center operator.

## 5 Conclusion

The role of the "forgotten" pseudo-momentum in the solution of the Landau problem as for uniform so in spatially non-uniform magnetic fields has been discussed in the series of the papers (see [11] and citing herein). Presented approach differs in that we placed particular emphasis on the gauge invariance of the procedure of the construction of the corresponding wave equations. It is common consensus in that the gauge invariance is one of the most fundamental symmetry properties of physics [55,56]. Thus, citing J. Schwinger [23], we follow his prescription that "a formally gauge invariant theory is ensured if one employs methods of solution that involve only gauge covariant quantities". In our paper we outlined the gauge invariant approach to the construction of the analog of guiding center operator in homogeneous/inhomogeneous magnetic fields. On the face of it, the presented approach is unnecessary and superfluous, as e.g. there exists common consensus that in the discussed "axial" problems (e.g., $\boldsymbol{B}(\boldsymbol{r}) = B_0 x \hat{\boldsymbol{z}}/L$) we must choose the wave function in the form $\Psi(\boldsymbol{r}) = \varphi(x) \exp(i\lambda y)$ simply relying on symmetry considerations. We argue that in accord with presented above considerations this is not so simple. Such naïve approach is valid only for the "proper" chosen gauge. The phase dependence of the wave function factor is determined by the eigenfunction of the additional time independent operator –pseudo-momentum, which in its turn is explicitly dependent upon the particular choice of the vector potential form. The steps to be taken to obtain the "proper" equations for the wave functions are as follows. First, we are free to choose any form of the vector potential fulfilling condition $rot\,\boldsymbol{A}(\boldsymbol{r}) = \boldsymbol{B}(\boldsymbol{r})$ appropriate to the considered magnetic field spatial distribution. Second, we define the conserved pseudo-momentum operator (or its component) dependent upon chosen gauge but nevertheless by the definition gauge invariant [57]. Fixing the phase of the exponential factor by imposing the restriction that the sought wave functions are simultaneously the eigenfunctions of the pseudo-momentum, we arrive at last to the gauge invariant equation. It is easy to verify that following these steps we always obtain the solutions which can be mapped in different gauges upon each other by traditional Weyl gauge transformation.

*Acknowledgements*

The work has been supported in part by the Ministry of Science and Higher Education of the Russian Federation under Project #FEUZ-2020-0054.